\documentclass[%
 reprint,
nofootinbib,
 amsmath,amssymb,
 aps,
 prab,
 floatfix,
]{revtex4-2}

\usepackage{graphicx}
\usepackage{dcolumn}
\usepackage{bm}
\usepackage{hyperref}
\usepackage{siunitx}

\newcommand{\s}{\enspace}

\begin{document}


\title{Identification of Magnetic Field Errors in Synchrotrons\\ based on Deep Lie Map Networks}

\author{Conrad Caliari${}^{1}$}
\author{Adrian Oeftiger${}^{2}$}
\author{Oliver Boine-Frankenheim${}^{1,2}$}%

\affiliation{%
 ${}^{1}$Institute for Accelerator Science and Electromagnetic Fields, Technische Universit\"at Darmstadt, Schlossgartenstr.\ 8, 64289 Darmstadt, Germany
}%

\affiliation{
${}^{2}$GSI Helmholtzzentrum f\"ur Schwerionenforschung GmbH, Planckstr.\ 1, 64291 Darmstadt, Germany
}%

\date{\today}

\begin{abstract}
Magnetic field errors pose a limitation in the performance of synchrotrons, as they excite non-systematic resonances, reduce dynamic aperture and may result in beam loss. Their effect can be compensated assuming knowledge of their location and strength. Established identification procedures are based on orbit response matrices or resonance driving terms. While they sequentially build a field error model for subsequent accelerator sections, a method detecting field errors in parallel could save valuable beam time. We introduce deep Lie map networks, which enable construction of an accelerator model including multipole components for the magnetic field errors by linking charged particle dynamics with machine learning methodology in a data-driven approach. Based on simulated beam-position-monitor readings for the example case of SIS18 at GSI, we demonstrate inference of location and strengths of gradient and sextupole errors for all accelerator sections in parallel. The obtained refined accelerator model may support setup of corrector magnets in operation to allow more precise control over tunes, chromaticities and resonance compensation.
\end{abstract}

\maketitle


\section{Introduction}
Synchrotron performance requirements necessitate detailed knowledge of magnetic fields errors present in the accelerator in order to minimize losses and maintain beam quality. Magnetic field errors cause beam loss by resonance excitation and demand compensation schemes if the working point cannot be freely changed. While particle tracking simulations are suited to predict and optimize corrector magnet settings with respect to beam loss, they depend on detailed knowledge of the magnetic field errors distributed along the beam line. Magnetic imperfections due to misalignments and fabrication errors are often present and their location and magnitude is essential for conclusive simulations. This work emphasizes that an accurate field error model can be obtained from a systematic comparison of simulation results to measurements. The magnetic field errors are recovered by minimizing discrepancies between predicted and measured motion of bunch centroids.

Existing approaches are based on measurements which assert the effect of steerer magnets or closed orbit bumps in a systematic scheme. Ref.~\cite{Safranek:LOCO} establishes the LOCO (linear optics form closed orbits) algorithm to model linear field errors. The orbit response matrix $A_{ij}$, i.e. the change in position at the i-th beam position monitor caused by a modification to the j-th corrector magnets deflection angle, is measured and compared to predictions of a computer model. Fitting the predicted to the measured orbit response matrix by variation of magnetic multipole components yields dipole, as well as normal and skew quadrupole field errors. Several methods have been proposed to obtain a non-linear magnetic field error model of a synchrotron by measurements. In \cite{Tomas:LocalResonanceTerms} the beam is excited by an ac dipole or a transversal kick in order to retrieve resonance driving terms. In \cite{Parfenova:NTRM} the authors propose to observe tune shifts induced by field errors in case the orbit is distorted globally. The effect of steerer magnets distributed along the synchrotron on measured tunes yields a response matrix, comparable to the orbit response matrix, and access to non-linear field errors. Order and resolution of the searched multipole components are limited by the resolution of the tune measurement, which relies on excitation of betatron oscillations by a kicker . These methods assume good knowledge of the linear field components.

Machine learning techniques yield the potential to model physical systems in a data-efficient way. This promises the identification of magnetic field errors without the time-consuming measurement of an orbit response matrix or installation of orbit bumps around the accelerator, but from few trajectories observed after excitation by a transversal kick. Physics-informed neural networks (PINN) have recently gained track in data-driven modelling of physical systems \cite{Raissi:PINNs,Cai:HeatTransferPINN}. They consist of an universal function approximator, the neural network, which is trained to reproduce measurements while obeying the physical laws governing the dynamics of the modelled system. This is achieved by minimizing a scalar loss function $\mathcal{L} = \mathcal{L}_\text{data} + \mathcal{L}_\text{reg}$, where $\mathcal{L}_\text{data}$ quantifies the discrepancy between prediction and measurement. The second term $\mathcal{L}_\text{reg}$ as a regularization term restricts the neural network to the space of possible solutions to the differential equations which express the considered laws of physics. A physics-informed neural network based on Taylor map layers has been successfully applied to orbit correction in \cite{Ivanov:PNN_beamDynamics}, where the symplecticity of Hamiltonian systems is enforced as a soft constraint by $\mathcal{L}_\text{reg}^\text{symp}$. The approach yields an effective model in form of transformations of particle coordinates between beam-position monitors.

This model works well for correction of closed orbit distortion due to dipole errors, also for application in simulations to the heavy-ion synchrotron SIS18 at GSI \cite{Caliari:Identification}. The Taylor map based PINN is capable of describing the phase advance between beam-position monitors, but after training predicted tunes show deviations of several percent. We observe a systematic failure to predict non-linear dynamics arising from sextupole errors. Since symplecticity is not strictly maintained during training by $\mathcal{L}_\text{reg}^\text{symp}$, the training results remain poor. This is possibly linked to failure modes related to the regularization term described in \cite{Krishnapriyan:PINNFailureModes}. 

As an alternative approach, we propose to replace the neural network by an accelerator model based on Lie algebra techniques and the thin-lens approximation \cite{Berz:IntroBeamPhysics}and embed it into the framework of machine learning. This model will be referred to as deep Lie map network (DLMN) in this work. In contrast to a PINN, this model choice does not involve a search for suited network structures, which is a non-trivial problem frequently tackled by a trial-and-error approach. By design, the DLMN represents a symplectic solution to the equations of motion, its degrees of freedom are given by magnetic multipole components. Hence, the DLMN approach enables a physical interpretation of the model and its degrees of freedom during any stage of the training process.

The large number of magnets constituting a synchrotron together with non-linear beam dynamics form a complex and high-dimensional optimization problem. We demonstrate the potential of the DLMN model trained by means of the ADAM \cite{Adam} algorithm to identify magnetic field errors in synchrotrons in simulations. Randomly distributed gradient and sextupole errors can be identified in the SIS18 synchrotron \cite{Ondreka:SIS18_recommissioning} in simulations, and tunes and chromaticities as well as resonance diagrams are reproduced in good agreement with the accelerator simulation providing training data.

The contribution is structured as follows: Section \ref{sec:Model} introduces the DLMN model, its training procedure is described in Section \ref{sec:TrainingProcedure} and simulation results for the SIS18 synchrotron are reported in Section \ref{sec:SimulationResults}. A conclusion is given in Section \ref{sec:Conclusion}.

\section{DLMN Model}
\label{sec:Model}
Crucial to the objective of creating an accurate representation of the accelerator is the chosen modelling approach. In this work, the modelling approach considers only drift spaces and transverse magnetic fields. The particle beam is reduced to a single particle representing its centroid. The equation of motion for single particle dynamics can be solved approximately in the framework of Hamiltonian dynamics \cite{Herr:Accelerator}. The thin-lens approximation consists of consecutive updates to position and momentum, known as drifts and kicks. The quality of the approximation depends on the order of the symplectic integrator, the arrangement of drifts and kicks, and symplectic integrators up to arbitrary order are known \cite{Yoshida:SymplecticIntegrators}. The particle tracking algorithms in MAD-X \cite{Grote:Mad-X} and SixTrackLib \cite{Schwinzerl:SixTrackLib} for instance, are based on this approach. 

Similar to layers of neurons forming neural networks, the accelerator model in thin-lens approximation consists of a concatenation of simple building blocks, the drifts and kicks. The transfer map of a single lattice element $\mathcal{M}_l$ arises from drifts $D$ and kicks $K (\vec{k})$,
\begin{equation}
    \mathcal{M}_l (\vec{k}) = K_i^l  \circ D_i^l \circ ... \circ K_1^l  \circ D_1^l \quad ,
\end{equation}
which depend on the multipole strengths $\vec{k}$ that characterize the present magnetic field. The transfer map between two locations in the lattice $\mathcal{M}_{l \to m}$ like beam-position monitors is given by their concatenation
\begin{equation}
    \mathcal{M}_{l \to m} = \mathcal{M}_m \circ ... \circ \mathcal{M}_{l+1} \circ \mathcal{M}_l \quad . 
\end{equation}
Analogous to layers in machine learning terminology, drifts and kicks represent elementary operations in terms of automatic differentiation. This enables its implementation in the framework of PINNs and allows to leverage existing tooling.

The model is capable of representing lattice magnets including linear fringe fields, but lacks rf cavities. Drift spaces are modelled without Taylor expansion of the square root and, thus, include non-linear effects like natural amplitude detuning. Chromatic detuning due to finite momentum spread of the beam and non-linear effects like amplitude detuning, cause motion of the beam centroid to deviate from motion of a single particle. The resolution of magnetic field errors correspondingly depends on transversal emittances as well as momentum spread of the beam. The effect of detuning on centroid motion is restricted by limiting the collection of training data to only a few turns after the beam is excited by a transversal kick. This is short compared to the synchrotron period and thus, we neglect rf cavities in the model. Collective effects like space charge, wakefields or electron clouds are neglected. An advanced implementation of the DLMN model could account for space charge, which can be included in terms of automatic differentiation as well \cite{Qiang:DifferentiableSpaceCharge}.

\section{Training Procedure}
\label{sec:TrainingProcedure}
Training describes the process of fitting the accelerator model to measurement data acquired by beam-position monitors. The optimal fit parameters reveal insights into the distribution of field errors since they represent magnetic multipole components. Apart from the model being subject to training itself, central to training are the training data, a metric quantifying the discrepancy between model predictions and training data, referred to as loss $\mathcal{L}$, and an optimization algorithm suited to minimize $\mathcal{L}$. In case of successful training, the model is capable of reproducing the measurements forming the training set and generalization beyond. Throughout the article we refer to the accelerator model being subject to training as the \textit{model}, whereas the source of training data, which is either a simulated or real machine, is referred to as \textit{accelerator}. The capability of the \textit{model} to predict correct trajectories from initial conditions not included into the training set is confirmed by additional data in a validation set.

\subsection{Training Data}
\label{sec:TrainingData}
The training set consists of measured centroid trajectories, which shall be reproduced by the \textit{model}. In order to predict the motion of the beam centroid an initial condition must be given as input to the \textit{model}. In machine experiments, an initial condition can be created by means of a kicker deflecting the beam from its equilibrium state. The kicker affects the beam in both planes and the  transversal momentum of the beam centroid is inferred from the kicker voltage. Additionally, the beam energy may be offset by slightly mismatching the rf frequency with respect to the revolution frequency for a given magnetic rigidity. A set of training data $\mathcal{T}$ is obtained by varying the kick strength and / or the rf frequency, while the beam  position monitors are used to observe the beam centroid motion.

Convergence speed is found to increase if training is performed in two stages. In the first stage, initial conditions used for training
\begin{equation}
    \mathcal{T}_1 = \{-\Delta p_x, \Delta p_x\} \times \{-\Delta p_y, \Delta p_y \}
\end{equation}
comprise horizontal $\Delta p_x$ resp.\ vertical $\Delta p_y$ excitation amplitude via the kicker. This allows a first estimate of gradient errors. In the second stage, off-momentum initial conditions are used for training,
\begin{equation}
    \mathcal{T}_2 = \{-\Delta p_x, \Delta p_x\} \times \{-\Delta p_y, \Delta p_y \} \times \{-\delta, \delta \}
\end{equation}
enabling identification of sextupole errors and chromaticities with high fidelity.

Deviations between single particle motion, as predicted by the \textit{model}, and centroid motion of realistic particle distributions grow over time due to chromatic and amplitude detuning. Hence, the number $M$ of turns shall be small. In addition, the computational complexity of tracking grows with $M$. A magnetic field error influences the centroid motion globally regardless of its location. It is therefore essential to observe the centroid motion for more than one turn to include the periodicity of a synchrotron. In case a single turn is used for training, we find the algorithm underestimates the relevance of field errors located close to the end of the turn, as they affect only few BPM readings downstream. This issue is mitigated by considering more than one turn, and we observed no further improvement for $M>3$ in the range of up to seven turns.

In case of SIS18, we find $M=3$ to be a good setting for the considered accelerator, which is short compared to the observation time necessary for a tune measurement required by alternative approaches.

\subsection{Loss}
In order to judge the quality of \textit{model} predictions they need to be compared to observations. A metric $\mathcal{L}(q(\vec{z}_0), \hat{q}(\vec{z}_0))$ called loss is introduced to quantify the discrepancy between \textit{model} output $\hat{q}(\vec{z}_0)$ and measurement $q(\vec{z_0})$.\\
This work makes use of a modified version of the mean-squared error (MSE) common to machine learning and regression. The loss
\begin{equation}
    \mathcal{L}(\{q, \hat{q}\}_{r \in R}) = \frac{1}{M} \sum_{r=1}^R \sum_{m=1}^M \sum_{n=1}^N \frac{\left( \vec{q}_{r,m,n} - \hat{\vec{q}}_{r,m,n} \right)^2}{\sigma_r^2} 
\end{equation}
compares predicted and measured centroid positions $\vec{q} = \begin{bmatrix} x, y\end{bmatrix}^T$ at discrete locations of $N$ beam position monitors over $M$ turns for $R$ initial conditions.
The normalization factor
\begin{equation}
    \sigma_r \left( \vec{z}_0^{(r)} \right) = \max \{ A_q \left( \vec{z}_0^{(r)} \right) \}_{q \in \{x,y\}}
\end{equation}
is given by the single-particle amplitude
\begin{equation}
    A_q = \sqrt{ 2 \beta_q J_q + D_q^2 \delta^2} \quad ,
\end{equation}
which depends on the initial condition $\vec{z}_0^{(r)} \equiv [p_x, p_y, \delta]$. Beta-functions $\beta_q$, dispersions $D_q$ and linearized actions $J_q$ are computed from the initial accelerator model. $\mathcal{L}$ is a positive-semidefinite function, and the case $\mathcal{L}=0$ indicates perfect agreement between \textit{model} predictions and data.

Since the loss $\mathcal{L}$ compares \textit{model} predictions to measurements, it depends on the magnetic multipole strengths $\vec{k}$ of the \textit{model}.
The optimal multipole strengths $\vec{k}^*$ satisfy for all $\vec{k} \in \mathcal{D}$: $\mathcal{L}(\vec{k}) \geq \mathcal{L}(\vec{k}^*)$ over some set of field strengths $\mathcal{D}$. This entails that the \textit{model} reproduces measured trajectories.
Thus, a comparison of the converged multipole strengths to those of the untrained initial model reveals magnetic field errors present in the accelerator.

\begin{figure}[t]
    \centering
    \includegraphics{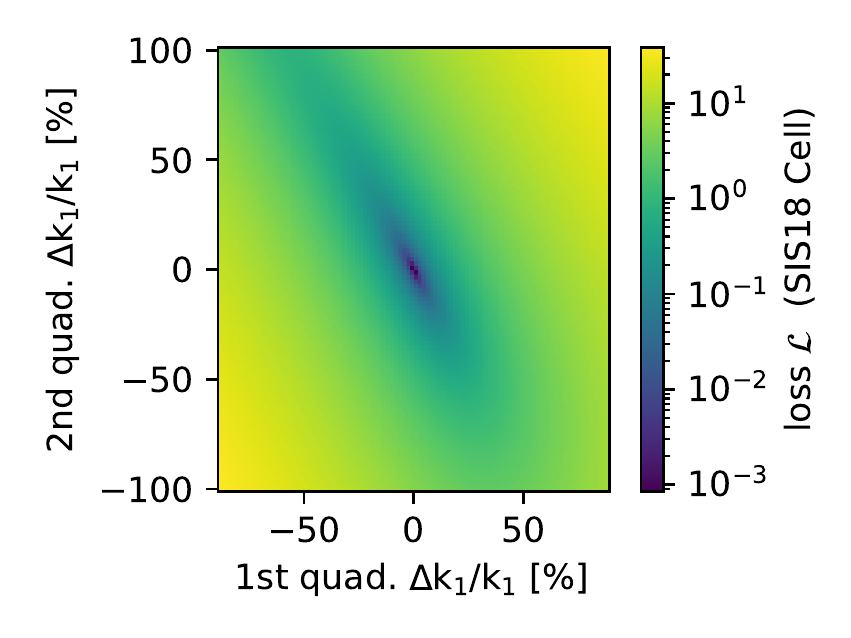}
    \caption{Loss function $\mathcal{L}$ of a SIS18 cell vs.\ gradient errors of first and second quadrupole. The training set consists of a single condition $\{p_x, p_y, \delta\} = \{10^{-3}, 10^{-3}, 0\}$}
    \label{fig:FODO_Loss}
\end{figure}

For a single FODO cell in thin-lens approximation, the eigenvalues of the Hessian of the loss $\mathcal{L}$ with respect to quadrupole strengths can be calculated analytically.
For not too large gradient errors, the Hessian is positive-definite and thus optimization of $\mathcal{L}$ poses a convex optimization problem, e.g.\ a unique extremum exists on $D$. In case of non-linear beam dynamics, e.g.\ non-linearities originating from truncation-free drifts and lattice sextupoles powered to correct chromaticity, this finding is not altered: A scan of the loss $\mathcal{L}$ as a function of \textit{model} quadrupole strengths shows a unique minimum in case the \textit{model} matches the quadrupole strengths of the accelerator, cf.\ Fig.~\ref{fig:FODO_Loss}. This emphasizes that minimization of $\mathcal{L}$ is a well-posed regression problem. The proposed method is then applied to nonlinear field components in  Section~\ref{sec:SimulationResults}.

\subsection{Optimization}
The DLMN model is trained by minimizing the loss over the training data set. Since the \textit{model} is differentiable, gradient-based optimization algorithms, which are established in various high-dimensional fit problems of machine learning, can be employed. In simulations, the ADAM \cite{Adam} algorithm outperformed options like plain gradient descent, Adagrad \cite{Adagrad} or Adadelta \cite{Adadelta}. The ADAM optimizer is capable of dealing with sparse gradients and parameters whose gradients differ in size by orders of magnitude. The derivatives of the loss with respect to \textit{model} parameters are obtained by automatic differentiation. 

Automatic differentiation \cite{Bartholomew:AutomaticDifferentiation} leverages that the \textit{model} consists of a concatenation of simple maps, the drift and kicks, which can be differentiated analytically in closed form. The derivatives of the whole \textit{model} are then calculated by exploitation of the chain rule, which allows to break down their calculation to a concatenation of the analytic derivatives of drift and kicks, similar to the concatenation of drifts and kicks yielding the particle tracking simulation in the first place. Since the scalar loss function is differentiated w.r.t.\ many multipole strengths characterizing each kick, we employ reverse-mode automatic differentiation, which is more efficient than forward-mode automatic differentiation in this case. In contrast to numerical differentiation based on finite differences, automatic differentiation is not prone to rounding errors and thus, noisy gradients. The derivation of an analytic expression for loss derivatives is infeasible because of expression swelling, which causes the number of terms to grow exponentially with the number of drifts and kicks.

The DLMN model as well as the training procedure are implemented in the Julia programming language \cite{Bezanson:Julia}. Automatic differentiation is used via the library \cite{Rackauckas:Zygote}, an implementation of the ADAM algorithm is taken from \cite{Innes:FluxML}. Additionally, the learning rate of the optimizer is decreased exponentially as a function of iterations over the training set.

Both learning rate, also known as step size, and its decay rate form two hyperparameters of the training procedure. A tree-structured Parzen estimator \cite{Bergstra:TPE} implemented in Ref.~\cite{Akiba:Optuna} is employed for their optimization. We find that the hyperparameters need rather limited tuning and optimal values are identified within few iterations of the Parzen estimator.

\section{Application to SIS18 in Simulations}
\label{sec:SimulationResults}

\begin{figure}[t]
    \centering
    \includegraphics[width=0.9\columnwidth]{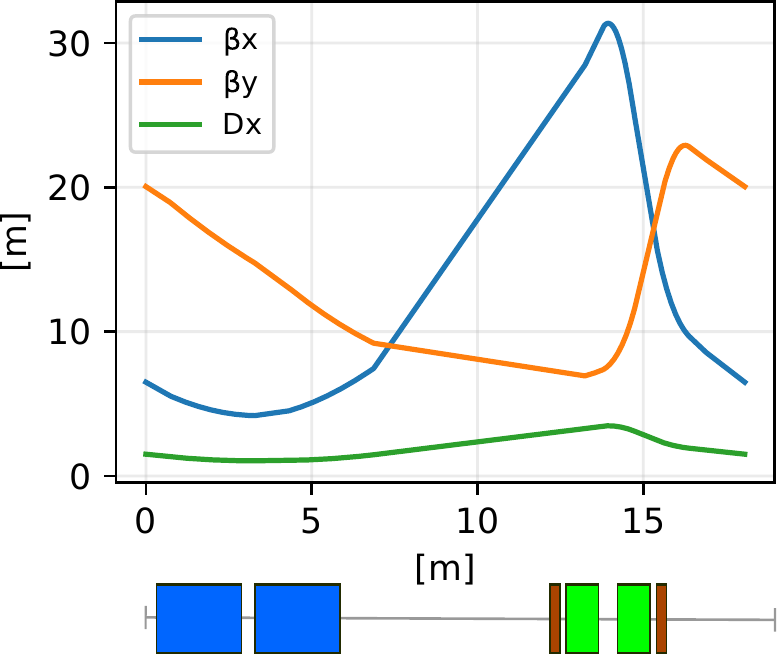}
    \caption{A \SI{18}{\meter} long cell of SIS18 drawn to scale. Bending magnets are shown in blue, quadrupoles in green and sextupoles in red.}
    \label{fig:SIS18CellLayout}
\end{figure}

The DLMN model is applied to the SIS18 at GSI in simulations. The SIS18 is a \SI{216}{\meter} long synchrotron designed to accelerate ions ranging from protons to uranium. It features twelve identical cells, where each cell hosts two bending magnets, two quadrupoles as well as two sextupole magnets, cf.\ Fig.~\ref{fig:SIS18CellLayout}. Training is performed on a detailed simulation of SIS18 based on the MAD-X / SixTrackLib codes, where the former is used for matching of tunes and chromaticities and the latter for 6D particle tracking. The simulation model consists of lattice magnets including linear fringe fields. Furthermore, the simulation includes an rf cavity enabling a bunched beam necessary for usage of the beam position monitors. The example beam used in this study consists of protons injected by the UNILAC \cite{Barth:Unilac}. The energy spread is determined by the injected micro-bunches \cite{Appel:Microbunch}, the chosen energy and the bunching factor. Here we assume an energy spread $\sigma_E / E=\num{1e-4}$ after acceleration, which can be achieved with an optimized ramp. The particle distribution within the beam is given by a matched 6D-Gaussian distribution.
Key beam parameters used in simulations can be found in Table~\ref{tab:BeamParameters}.

The simulated synchrotron features various combinations of gradient and sextupole errors, and provides beam-position monitor readings of the beam centroid position as outputs.
The recovery of field errors hidden in the accelerator simulation is limited by the \textit{models} approximation of the beam by a single particle. Thus, the resolution of gradient and sextupole errors is evaluated in dependence of transverse emittance and energy spread of the beam.

\begin{table}[tb]
    \centering
    \caption{Properties of SIS18 and Key Beam Parameters \cite{Barth:Unilac,Franczak:SIS18_parameterList}.}

    \begin{tabular*}{\linewidth}{l@{\extracolsep{\fill}}c}
        \hline
        \hline
        Parameter & Value \\ \hline
        Circumference   &   \SI{216}{\meter}    \\
        Momentum compaction $\alpha_C$  &   \num{3.4e-2}    \\
        Transition energy $\gamma_T$  &   5.5    \\
        Synchrotron tune $Q_s$   &   $>$ \num{100} turns \\
        Betatron tunes $Q_x$, $Q_y$  &   4.2, 3.4   \\
        Natural (absolute) chromaticity $\xi_x^\text{(nat)}$, $\xi_y^\text{(nat)}$    &   -6.43 / -4.89   \\
        Magnetic rigidity $(B \rho)^\text{max}$  &   \SI{18.5}{\tesla\meter}   \\
        Ion &   proton  \\
         Energy $E$  &   \SI{5}{\giga\electronvolt}  \\
         Energy spread $\sigma_E / E$  &   \num{1e-4} \\
         Transverse emittances $\epsilon_\text{norm}^\text{4-rms}$  &  \SI{0.9}{\micro\meter} \\
         \hline
         \hline
    \end{tabular*}
    \label{tab:BeamParameters}
\end{table}

In Subsection \ref{ref:Resolution} the possible resolution of gradient and sextupole errors in dependence of beam parameters is discussed. Subsection \ref{ref:OrbitDistortion} covers the case in which the \textit{model} lacks a degree of freedom at the location of a field error. The simultaneous identification of a set of distributed field errors is presented in Subsection \ref{ref:RandomErrors}. Physical plausibility of the \textit{model} predictions is underpinned by correct prediction of tunes and chromaticities.
\subsection{Resolution}
\label{ref:Resolution}
\begin{figure*}[t]
    \centering
    \includegraphics{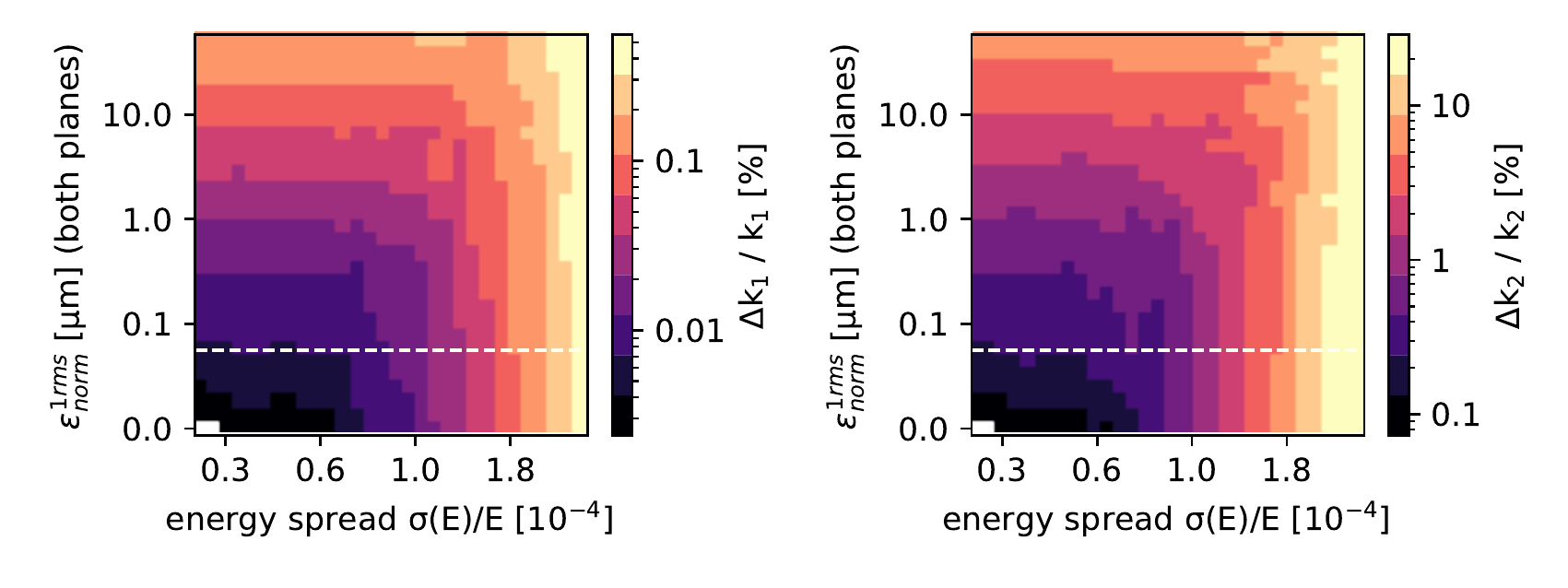}
    \caption{Resolution of gradient (left) and sextupole (right) errors in dependence of beam transverse emittance and energy spread. A gradient $k_1L=$\SI{5.2e-3}{\per\meter} and a sextupole $k_2L=$\SI{1.6e-2}{\per\meter^2} field error are introduced to the lattice. The dashed line marks the normalized 1-rms emittance of the UNILAC proton beam.}
    \label{fig:EmittanceVsSige}
\end{figure*}

The resolution of magnetic multipole components is limited by the approximation of the beam centroid by a single particle. Due to adiabatic damping, the transverse beam size is smallest at high energy. Thus, non-linear effects such as amplitude detuning become less influential. To benefit from this effect, training data is collected at flat-top energy. A single gradient error $k_1L=\SI{5.2e-3}{\meter^{-1}}$ together with a single sextupole error $k_2L=\SI{1.6e-2}{\meter^{-2}}$ are introduced to the accelerator, 
causing a shift in tune $\Delta Q_x / Q_x \approx\num{1.2e-3}$ and chromaticity $\Delta \xi_x / \xi_x \approx\num{1.7e-2}$. Training is performed for different transverse emittances as well as energy spreads and the achieved resolution of field errors is quantified by the discrepancy 
\begin{equation}
    D_i = \left| \frac{k_i^\text{acc} - k_i^\text{model}}{k_i^\text{acc}} \right|
\end{equation}
 between the \textit{models} multipole strength and its actual counterpart in the accelerator, which is displayed in Fig.~\ref{fig:EmittanceVsSige}. The discrepancy in both gradient and sextupole strengths is determined primarily by the beam energy spread. Additionally, the discrepancy in sextupole strengths grows beyond \SI{10}{\percent} in case the normalized 1-rms emittance exceeds \SI{10}{\micro\meter}. The beam size of the proton beam is suited for resolving sextupole components in the order of magnitude $k_2L \approx\SI{e-2}{\meter^{-2}}$ with a discrepancy of $D_2 < \SI{10}{\percent}$. Therefore, it is suited to identify undocumented sextupole contributions related to the main dipoles in SIS18, but resolution is reduced significantly in case of heavy-ion beams featuring larger transverse emittances.

\subsection{Orbit Distortion}
\label{ref:OrbitDistortion}
In addition to field errors, another source of deviations between accelerator and \textit{model} is distortion of the closed orbit. Besides moving the center of betatron oscillations, a displacement $d$ of the closed orbit with respect to the geometric centre of a magnet induces multipole components of lower orders. In case the magnet is a $2(n+1)$-pole the dominant feed-down contribution acts like a $(2n)$-pole, i.e.\ an orbit distortion inside a sextupole field $k_2$ induces a gradient component $k_1^\text{sext}$,
\begin{align}
\begin{split}
    \Delta p_x &= \frac{k_2 L}{2} (x + d)^2 \\
    &= \frac{k_2 L}{2} x^2 + \underbrace{k_2 L\, d }_{k_1^\text{sext}} \cdot x + \frac{k_2 L}{2} d^2 \quad 
\end{split}
\end{align}
This effective $k_1^\text{sext}$ yields a corresponding tune shift. Therefore, training in presence of feed-down yields an effective model, whose multipole components may differ from those given a series expansion around a magnets geometric center.

In order to train the \textit{model} in presence of orbit distortions, BPM readings are aligned to zero mean. The remaining deviations originate from feed-down. We investigate the effect of a closed orbit bump inside a single sextupole, for the scenario that sextupoles are used to correct chromaticity to zero in both transverse planes in SIS18. The simulated accelerator features a 6D-Gaussian beam profile according to beam parameters found in Table~\ref{tab:BeamParameters}. The degrees of freedom of the \textit{model} comprise the focusing strengths of the quadrupoles, which will be adjusted during the training process. The orbit bump leads to an effective gradient error at the location of the sextupole, which cannot be resolved by training because the \textit{model} lacks a gradient degree of freedom at the sextupole location.

\begin{figure}[t]
    \centering
    \includegraphics{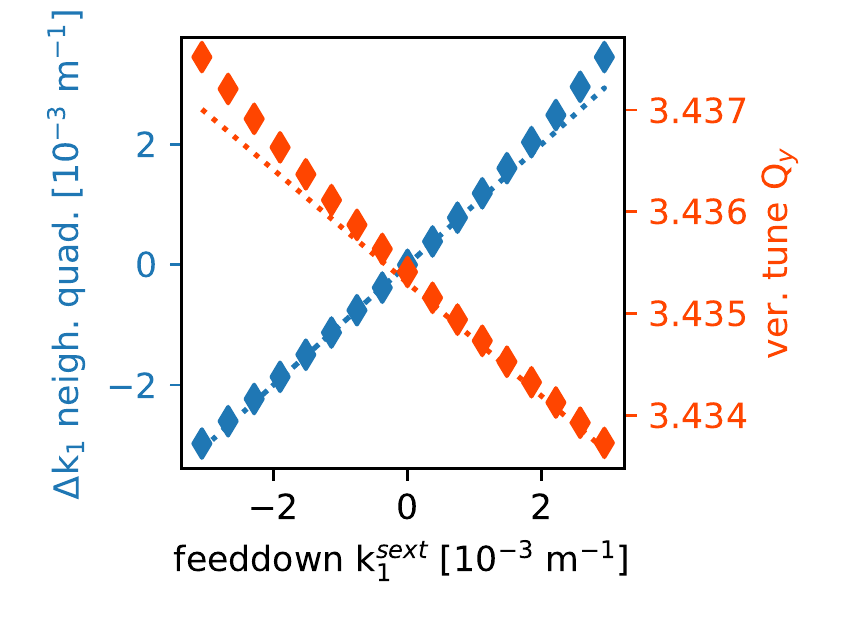}
    \caption{Training results for a closed orbit distortion scenario. The final quadrupole deviation of the closest neighboring quadrupole after training is compared to the gradient feed-down induced by the orbit bump. The vertical tune predicted by the \textit{model} is compared to the actual accelerator tune (dotted).}
    \label{fig:OrbitBumpResults}
\end{figure}

Training is capable to predict global properties, e.g. the tunes in both planes. As shown in Fig.~\ref{fig:OrbitBumpResults}, the tune shift induced by the orbit excursion is resolved accurately. We observe training to adjust the focusing strength of the closest located neighboring quadrupole, whereas all other focusing strengths as well the sextupole strengths are not altered by the algorithm. 

In contrast to all other quadrupole strengths, the strength of the neighboring quadrupole does incorporate the gradient error induced by the orbit bump in the sextupole. Training is observed to adjust the degree of freedom closest to the error and thus, localization of the cell hosting the error is possible. A large orbit bump induces an additional dipole error due to feed-down. Since the focusing strengths as degrees of freedom cannot reproduce the closed orbit, resolution worsens for large bump excursions.

The results emphasize the DLMN is capable of identifying gradient and sextupole errors in presence of small dipole errors. As a conclusion, the collection of training data ought to be preceded by a closed orbit correction to identify field errors with good resolution. 

\subsection{Random Field Errors}
\label{ref:RandomErrors}
Besides of systematic field errors, also random contributions due to fabrication errors and misalignments are likely to be distributed across the accelerator. During operation of SIS18, measurements of global properties like tunes and chromaticities differ from predictions by the existing accelerator model. This discrepancy is large especially in the case of chromaticities and depends on the excitation current of the dipole magnets. Therefore, it is of interest to investigate the applicability of the DLMN model to quantify sextupole components present in the accelerator ring.

\begin{figure*}[t]
    \centering
    \includegraphics{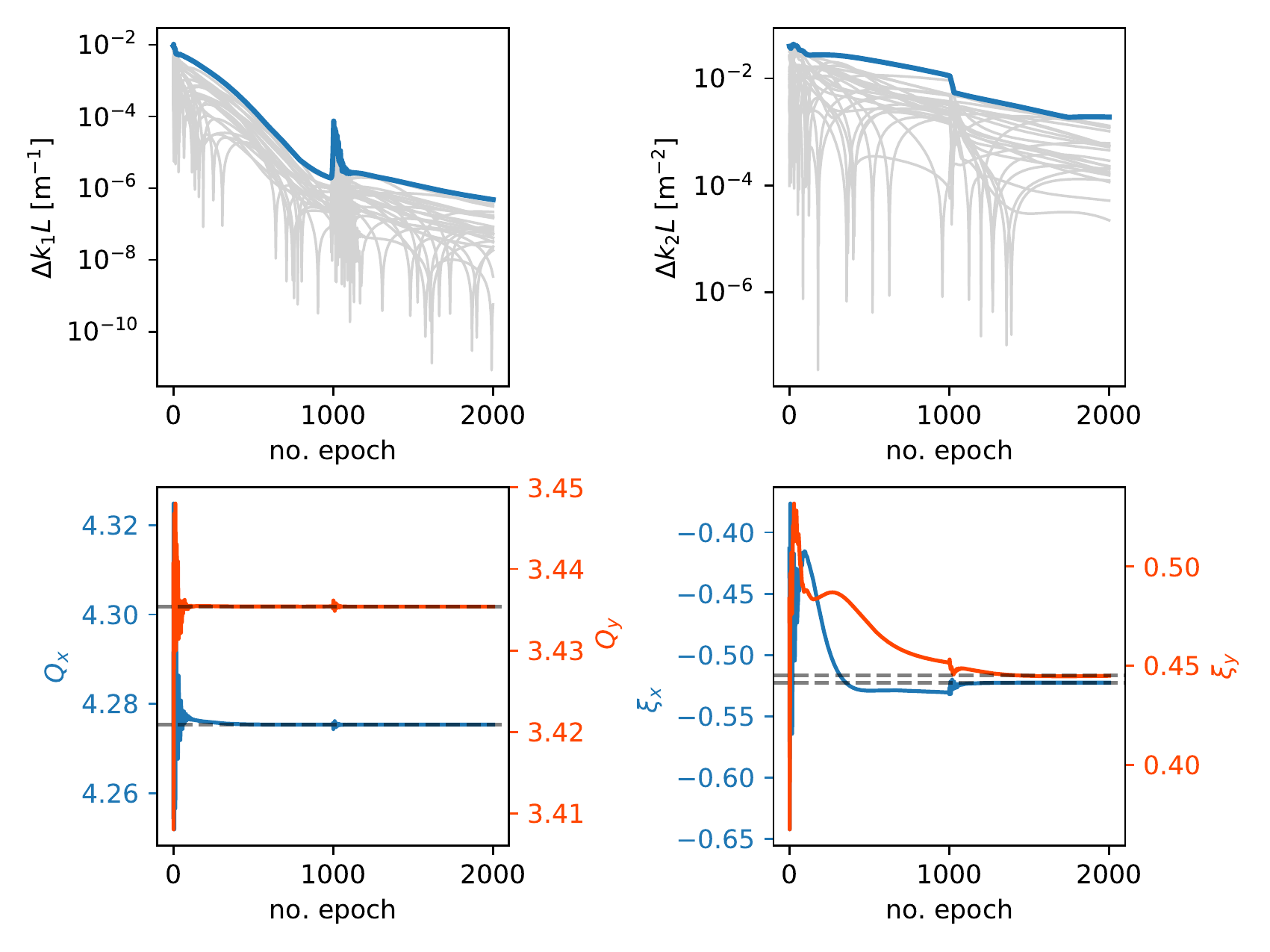}
    \caption{Training results in case of normal distributed gradient and sextupole errors. Top: The maximum deviation between gradient and sextupole strengths of \textit{model} and accelerator during training is shown in blue, gray lines represent individual multipole strengths. Bottom: Tunes $Q$ and chromaticities $\xi$ in both planes converge against those present in the accelerator, denoted by dashed lines. }
    \label{fig:MultipoleEvolution}
\end{figure*}

Random gradient and sextupole errors are added to the 24 main quadrupoles and 24 bending magnets of SIS18 in the simulation model. The error multipole strengths are sampled from a normal distribution with standard deviation $\sigma_\text{quad}$ and $\sigma_\text{sext}$ for gradient and sextupole errors, respectively. The magnitude is chosen such that each error perturbs betatron tune $Q$ and (absolute) chromaticity $\xi$ by
\begin{equation}
    \frac{\Delta Q_x}{Q_x} \left( \sigma_\text{quad} \right) = 10^{-3} , \s \frac{\Delta \xi_x}{\xi_{x,\text{nat}}} \left( \sigma_\text{sext} \right) = 8 \cdot 10^{-2} \s  , \notag
\end{equation}
likewise for the vertical plane. Similar to the scenario \textbf{B}, the beam used in the simulated accelerator follows a 6D-Gaussian particle distribution according to the emittances and the energy spread given in Table~\ref{tab:BeamParameters}. 

The DLMN model is tasked to identify normal distributed gradient and sextupole errors. Its degrees of freedom comprise sextupole strengths of the main dipoles and gradient strengths of the lattice quadrupoles. Training is capable of successfully minimizing the loss over the training set. Simultaneously, the discrepancy in multipole strength is significantly decreased for quadrupole as well as sextupole strengths. The switch of training sets to incorporate off-momentum trajectories into training is performed at epoch no.~\num{1000}, cf.\ Section \ref{sec:TrainingData}. This changes the magnitude of the calculated loss derivatives w.r.t.\ to the multipole strengths of the \textit{model}, which causes a peak in multipole deviations as the ADAM algorithm needs to adapt its internal step size. At each epoch, loss derivatives are calculated for each trajectory of the training set and the gradient descent takes a single step in the direction of the average gradient. Training on off-momentum trajectories, where $\delta=\num{5e-3}$, enables improved resolution of sextupoles, cf.\ Fig.~\ref{fig:MultipoleEvolution}. Observation of off-momentum trajectories is therefore essential to model sextupole components.

The evolution of tunes and chromaticities predicted by the DLMN model converge in both planes against their counterparts present in the accelerator simulation that generated the training data in the first place. The resolution of tunes exceeds typical measurement uncertainties of these quantities.

The DLMN model is found to be capable of predicting the magnitude of distributed gradient and sextupole errors present in SIS18 in simulations. 
The field errors are correctly identified for an accelerator setup both at natural chromaticity as well as for corrected chromaticity, $\xi_{x,y}\rightarrow 0$, where strong systematic sextupole fields are present in the lattice sextupoles.
The field errors identified during training can potentially explain observed discrepancies in tune and chromaticity in real accelerators. 

The training works just as well for other betatron tunes than the indicated SIS18 working point. In general, these random field errors drive non-systematic betatron resonances. In a dedicated study, the betatron tune has been varied scanning through a regular non-systematic sextupole resonance, cf.\ Appendix~\ref{appendix:ResolutionVsWorkingPoint}. As a result of the study, the resolution of the identified gradient and sextupole errors was found to be rather independent of the nearby resonance.

Therefore, the trained DLMN model can be applied to support operations for precise control of tunes and chromaticities, as well as resonance compensation.

\section{Conclusion}
\label{sec:Conclusion}
In order to identify magnetic field errors, this work combines conventional modelling approaches in beam dynamics with training techniques designed for artificial neural networks.
The proposed Deep Lie Map Network (DLMN) model enables identification of field errors based on observations of beam centroid motion by means of beam position monitors.
This data-driven modelling approach yields an effective model of the accelerator, which encapsulates location and magnitude of magnetic field errors. It can therefore be used to compute resonance diagrams and driving terms. In contrast to methods like the LOCO algorithm \cite{Safranek:LOCO}, the non-linear tune response matrix \cite{Parfenova:NTRM} or measurement of the resonance driving terms \cite{Tomas:LocalResonanceTerms}, the proposed method does not require the time-consuming systematic installation of closed orbit bumps around the synchrotron. The trained DLMN model predicts tunes and chromaticities in good agreement with the accelerator being subject to training. In the simulated example case of SIS18, the training procedure has been demonstrated to quantify gradient and sextupole errors. The effects of residual orbit distortions and decoherence on the resolution of field errors is analyzed by parameter scans. We conclude that DLMN may be applied to real synchrotrons, but the collection of training data must be preceded by a closed orbit correction. In principle, the developed training procedure can be applied to higher-order field errors like octupoles.

In contrast to a physics-informed neural network \cite{Ivanov:PNN_beamDynamics}, the DLMN approach inherently incorporates the symplectic structure of beam dynamics and is guaranteed to be a valid solution to the equations of motion.
The DLMN model parameters are physically meaningful magnetic multipole components and can, therefore, be interpreted at any stage of the training procedure. This warrants further use of the trained effective model in established tools and (tracking) codes of accelerator physics such as, for instance, MAD-X and SixTrackLib.
When modelling large accelerator rings, the present approach in thin-lens approximation may require a larger amount of concatenated drifts and kicks to obtain highly resolved field errors. In order to reduce the computing time in the context of automatic differentiation as required for the gradient-descent training algorithm, further research could refine the developed Lie map network by modelling thick elements based on the Truncated Power Series Algebra technique.

DLMN model training yields the potential to reduce the need for beam time dedicated to identify unknown magnetic field errors and establish an effective machine model, which may increase availability and performance of synchrotrons. The small size of the required training data set facilitates short time windows of data collection and, thus, monitoring of field errors throughout the year. The trained effective machine model may serve to support precise control of betatron tunes, chromaticities and resonance compensation.

\section*{Acknowledgements}
The authors thank Simon Hirl\"ander and Sabrina Appel for valuable discussions on hyperparameter optimization and modelling.


\appendix

\section{Dependence of Field Error Resolution on Working Point}
\label{appendix:ResolutionVsWorkingPoint}
\begin{figure*}[ht]
    \centering
    \includegraphics{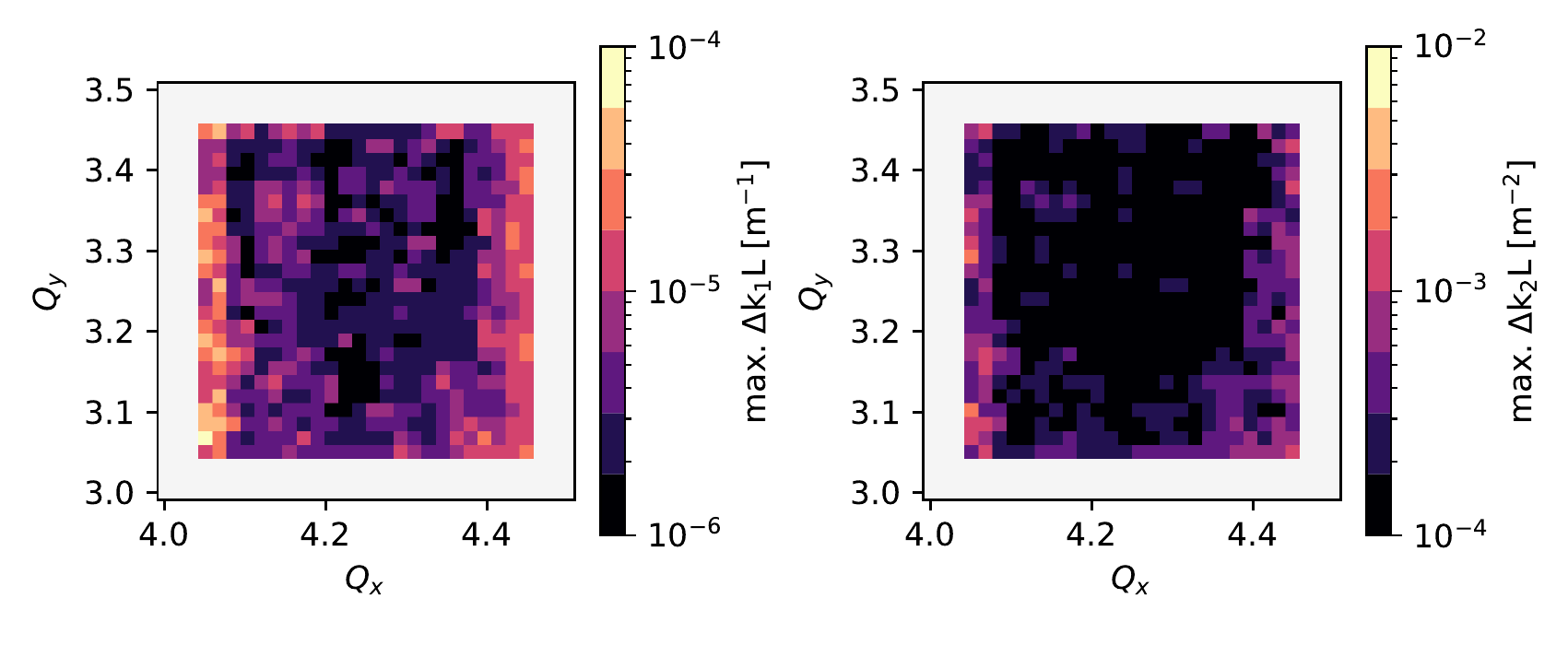}
    \caption{Dependence of the resolution of field errors on the working point. The plots show the maximum deviation in gradient (left) and sextupole (right) degrees of freedom after training. The results are averaged over seven different field error samples.}
    \label{fig:ResolutionVsWorkingPoint}
\end{figure*}

The dependence of the resolution of normal-distributed gradient and sextupole errors on the chosen working point is investigated. The training procedure and the simulated accelerator being source of the training data are set up similarly to the scenario depicted in Section \ref{ref:RandomErrors}. For each scan, the sample of field errors drawn from a normal distribution is kept constant and the maximum deviation $\Delta k_1L$ in quadrupole and $\Delta k_2L$ in sextupole components after training is recorded. The scan is repeated for different field error samples and the averaged results are shown in Fig.~\ref{fig:ResolutionVsWorkingPoint}.

The observed final deviations can be compared to results obtained in Section~\ref{ref:RandomErrors}, cf.\ Fig.~\ref{fig:MultipoleEvolution} and indicate the resolution of field errors as well as the prediction of correct tunes and chromaticities is possible independently of the working point. We find the resolution of sextupole errors is not affected by third-order resonances. 

\bibliography{Draft}

\begin{thebibliography}{27}%
\makeatletter
\providecommand \@ifxundefined [1]{%
 \@ifx{#1\undefined}
}%
\providecommand \@ifnum [1]{%
 \ifnum #1\expandafter \@firstoftwo
 \else \expandafter \@secondoftwo
 \fi
}%
\providecommand \@ifx [1]{%
 \ifx #1\expandafter \@firstoftwo
 \else \expandafter \@secondoftwo
 \fi
}%
\providecommand \natexlab [1]{#1}%
\providecommand \enquote  [1]{``#1''}%
\providecommand \bibnamefont  [1]{#1}%
\providecommand \bibfnamefont [1]{#1}%
\providecommand \citenamefont [1]{#1}%
\providecommand \href@noop [0]{\@secondoftwo}%
\providecommand \href [0]{\begingroup \@sanitize@url \@href}%
\providecommand \@href[1]{\@@startlink{#1}\@@href}%
\providecommand \@@href[1]{\endgroup#1\@@endlink}%
\providecommand \@sanitize@url [0]{\catcode `\\12\catcode `\$12\catcode
  `\&12\catcode `\#12\catcode `\^12\catcode `\_12\catcode `\%12\relax}%
\providecommand \@@startlink[1]{}%
\providecommand \@@endlink[0]{}%
\providecommand \url  [0]{\begingroup\@sanitize@url \@url }%
\providecommand \@url [1]{\endgroup\@href {#1}{\urlprefix }}%
\providecommand \urlprefix  [0]{URL }%
\providecommand \Eprint [0]{\href }%
\providecommand \doibase [0]{https://doi.org/}%
\providecommand \selectlanguage [0]{\@gobble}%
\providecommand \bibinfo  [0]{\@secondoftwo}%
\providecommand \bibfield  [0]{\@secondoftwo}%
\providecommand \translation [1]{[#1]}%
\providecommand \BibitemOpen [0]{}%
\providecommand \bibitemStop [0]{}%
\providecommand \bibitemNoStop [0]{.\EOS\space}%
\providecommand \EOS [0]{\spacefactor3000\relax}%
\providecommand \BibitemShut  [1]{\csname bibitem#1\endcsname}%
\let\auto@bib@innerbib\@empty
\bibitem [{\citenamefont {Safranek}(1997)}]{Safranek:LOCO}%
  \BibitemOpen
  \bibfield  {author} {\bibinfo {author} {\bibfnamefont {J.}~\bibnamefont
  {Safranek}},\ }\bibfield  {title} {\bibinfo {title} {{Experimental
  determination of storage ring optics using orbit response measurements}},\
  }\href {https://doi.org/10.1016/S0168-9002(97)00309-4} {\bibfield  {journal}
  {\bibinfo  {journal} {Nucl. Instrum. Meth. A}\ }\textbf {\bibinfo {volume}
  {388}},\ \bibinfo {pages} {27} (\bibinfo {year} {1997})}\BibitemShut
  {NoStop}%
\bibitem [{\citenamefont {Tom{\'a}s}\ \emph {et~al.}(2005)\citenamefont
  {Tom{\'a}s}, \citenamefont {Bai}, \citenamefont {Calaga}, \citenamefont
  {Fischer}, \citenamefont {Franchi},\ and\ \citenamefont
  {Rumolo}}]{Tomas:LocalResonanceTerms}%
  \BibitemOpen
  \bibfield  {author} {\bibinfo {author} {\bibfnamefont {R.}~\bibnamefont
  {Tom{\'a}s}}, \bibinfo {author} {\bibfnamefont {M.}~\bibnamefont {Bai}},
  \bibinfo {author} {\bibfnamefont {R.}~\bibnamefont {Calaga}}, \bibinfo
  {author} {\bibfnamefont {W.}~\bibnamefont {Fischer}}, \bibinfo {author}
  {\bibfnamefont {A.}~\bibnamefont {Franchi}},\ and\ \bibinfo {author}
  {\bibfnamefont {G.}~\bibnamefont {Rumolo}},\ }\bibfield  {title} {\bibinfo
  {title} {Measurement of global and local resonance terms},\ }\href@noop {}
  {\bibfield  {journal} {\bibinfo  {journal} {Physical Review Special
  Topics-Accelerators and Beams}\ }\textbf {\bibinfo {volume} {8}},\ \bibinfo
  {pages} {024001} (\bibinfo {year} {2005})}\BibitemShut {NoStop}%
\bibitem [{\citenamefont {Parfenova}\ and\ \citenamefont
  {Franchetti}(2011)}]{Parfenova:NTRM}%
  \BibitemOpen
  \bibfield  {author} {\bibinfo {author} {\bibfnamefont {A.}~\bibnamefont
  {Parfenova}}\ and\ \bibinfo {author} {\bibfnamefont {G.}~\bibnamefont
  {Franchetti}},\ }\bibfield  {title} {\bibinfo {title} {{Experimental
  benchmarking of nonlinear tune response matrix with several controlled
  sextupolar errors}},\ }\href {https://doi.org/10.1016/j.nima.2011.03.051}
  {\bibfield  {journal} {\bibinfo  {journal} {Nucl. Instrum. Meth. A}\ }\textbf
  {\bibinfo {volume} {646}},\ \bibinfo {pages} {7} (\bibinfo {year}
  {2011})}\BibitemShut {NoStop}%
\bibitem [{\citenamefont {Raissi}\ \emph {et~al.}(2019)\citenamefont {Raissi},
  \citenamefont {Perdikaris},\ and\ \citenamefont
  {Karniadakis}}]{Raissi:PINNs}%
  \BibitemOpen
  \bibfield  {author} {\bibinfo {author} {\bibfnamefont {M.}~\bibnamefont
  {Raissi}}, \bibinfo {author} {\bibfnamefont {P.}~\bibnamefont {Perdikaris}},\
  and\ \bibinfo {author} {\bibfnamefont {G.~E.}\ \bibnamefont {Karniadakis}},\
  }\bibfield  {title} {\bibinfo {title} {Physics-informed neural networks: A
  deep learning framework for solving forward and inverse problems involving
  nonlinear partial differential equations},\ }\href@noop {} {\bibfield
  {journal} {\bibinfo  {journal} {Journal of Computational physics}\ }\textbf
  {\bibinfo {volume} {378}},\ \bibinfo {pages} {686} (\bibinfo {year}
  {2019})}\BibitemShut {NoStop}%
\bibitem [{\citenamefont {Cai}\ \emph {et~al.}(2021)\citenamefont {Cai},
  \citenamefont {Wang}, \citenamefont {Wang}, \citenamefont {Perdikaris},\ and\
  \citenamefont {Karniadakis}}]{Cai:HeatTransferPINN}%
  \BibitemOpen
  \bibfield  {author} {\bibinfo {author} {\bibfnamefont {S.}~\bibnamefont
  {Cai}}, \bibinfo {author} {\bibfnamefont {Z.}~\bibnamefont {Wang}}, \bibinfo
  {author} {\bibfnamefont {S.}~\bibnamefont {Wang}}, \bibinfo {author}
  {\bibfnamefont {P.}~\bibnamefont {Perdikaris}},\ and\ \bibinfo {author}
  {\bibfnamefont {G.~E.}\ \bibnamefont {Karniadakis}},\ }\bibfield  {title}
  {\bibinfo {title} {Physics-informed neural networks for heat transfer
  problems},\ }\href@noop {} {\bibfield  {journal} {\bibinfo  {journal}
  {Journal of Heat Transfer}\ }\textbf {\bibinfo {volume} {143}} (\bibinfo
  {year} {2021})}\BibitemShut {NoStop}%
\bibitem [{\citenamefont {Ivanov}\ and\ \citenamefont
  {Agapov}(2020)}]{Ivanov:PNN_beamDynamics}%
  \BibitemOpen
  \bibfield  {author} {\bibinfo {author} {\bibfnamefont {A.}~\bibnamefont
  {Ivanov}}\ and\ \bibinfo {author} {\bibfnamefont {I.}~\bibnamefont
  {Agapov}},\ }\bibfield  {title} {\bibinfo {title} {{Physics-Based Deep Neural
  Networks for Beam Dynamics in Charged Particle Accelerators}},\ }\href
  {https://doi.org/10.1103/PhysRevAccelBeams.23.074601} {\bibfield  {journal}
  {\bibinfo  {journal} {Phys. Rev. Accel. Beams}\ }\textbf {\bibinfo {volume}
  {23}},\ \bibinfo {pages} {074601} (\bibinfo {year} {2020})},\ \Eprint
  {https://arxiv.org/abs/2007.03555} {arXiv:2007.03555 [cs.NE]} \BibitemShut
  {NoStop}%
\bibitem [{\citenamefont {Caliari}(2021)}]{Caliari:Identification}%
  \BibitemOpen
  \bibfield  {author} {\bibinfo {author} {\bibfnamefont {C.}~\bibnamefont
  {Caliari}},\ }\emph {\bibinfo {title} {Identification of Field Errors with
  Machine Learning Techniques}},\ \href@noop {} {Master's thesis},\ \bibinfo
  {school} {TU Darmstadt} (\bibinfo {year} {2021})\BibitemShut {NoStop}%
\bibitem [{\citenamefont {Krishnapriyan}\ \emph {et~al.}(2021)\citenamefont
  {Krishnapriyan}, \citenamefont {Gholami}, \citenamefont {Zhe}, \citenamefont
  {Kirby},\ and\ \citenamefont {Mahoney}}]{Krishnapriyan:PINNFailureModes}%
  \BibitemOpen
  \bibfield  {author} {\bibinfo {author} {\bibfnamefont {A.}~\bibnamefont
  {Krishnapriyan}}, \bibinfo {author} {\bibfnamefont {A.}~\bibnamefont
  {Gholami}}, \bibinfo {author} {\bibfnamefont {S.}~\bibnamefont {Zhe}},
  \bibinfo {author} {\bibfnamefont {R.}~\bibnamefont {Kirby}},\ and\ \bibinfo
  {author} {\bibfnamefont {M.~W.}\ \bibnamefont {Mahoney}},\ }\bibfield
  {title} {\bibinfo {title} {Characterizing possible failure modes in
  physics-informed neural networks},\ }in\ \href
  {https://proceedings.neurips.cc/paper/2021/file/df438e5206f31600e6ae4af72f2725f1-Paper.pdf}
  {\emph {\bibinfo {booktitle} {Advances in Neural Information Processing
  Systems}}},\ Vol.~\bibinfo {volume} {34},\ \bibinfo {editor} {edited by\
  \bibinfo {editor} {\bibfnamefont {M.}~\bibnamefont {Ranzato}}, \bibinfo
  {editor} {\bibfnamefont {A.}~\bibnamefont {Beygelzimer}}, \bibinfo {editor}
  {\bibfnamefont {Y.}~\bibnamefont {Dauphin}}, \bibinfo {editor} {\bibfnamefont
  {P.}~\bibnamefont {Liang}},\ and\ \bibinfo {editor} {\bibfnamefont {J.~W.}\
  \bibnamefont {Vaughan}}}\ (\bibinfo  {publisher} {Curran Associates, Inc.},\
  \bibinfo {year} {2021})\ pp.\ \bibinfo {pages} {26548--26560}\BibitemShut
  {NoStop}%
\bibitem [{\citenamefont {Berz}\ \emph {et~al.}(2015)\citenamefont {Berz},
  \citenamefont {Makino},\ and\ \citenamefont {Wan}}]{Berz:IntroBeamPhysics}%
  \BibitemOpen
  \bibfield  {author} {\bibinfo {author} {\bibfnamefont {M.}~\bibnamefont
  {Berz}}, \bibinfo {author} {\bibfnamefont {K.}~\bibnamefont {Makino}},\ and\
  \bibinfo {author} {\bibfnamefont {W.}~\bibnamefont {Wan}},\ }\href
  {https://doi.org/10.1201/b12074} {\emph {\bibinfo {title} {An {Introduction}
  to {Beam} {Physics}}}}\ (\bibinfo  {publisher} {Taylor \& Francis},\ \bibinfo
  {year} {2015})\ \bibinfo {note} {accepted: 2021-10-11T14:23:25Z}\BibitemShut
  {NoStop}%
\bibitem [{\citenamefont {Kingma}\ and\ \citenamefont {Ba}(2014)}]{Adam}%
  \BibitemOpen
  \bibfield  {author} {\bibinfo {author} {\bibfnamefont {D.~P.}\ \bibnamefont
  {Kingma}}\ and\ \bibinfo {author} {\bibfnamefont {J.}~\bibnamefont {Ba}},\
  }\href@noop {} {\bibinfo {title} {Adam: A method for stochastic
  optimization}} (\bibinfo {year} {2014}),\ \Eprint
  {https://arxiv.org/abs/1412.6980} {arXiv:1412.6980 [cs.LG]} \BibitemShut
  {NoStop}%
\bibitem [{\citenamefont {Ondreka}\ \emph {et~al.}(2019)\citenamefont
  {Ondreka}, \citenamefont {Dimopoulou}, \citenamefont {Hüther}, \citenamefont
  {Liebermann}, \citenamefont {Stadlmann},\ and\ \citenamefont
  {Steinhagen}}]{Ondreka:SIS18_recommissioning}%
  \BibitemOpen
  \bibfield  {author} {\bibinfo {author} {\bibfnamefont {D.}~\bibnamefont
  {Ondreka}}, \bibinfo {author} {\bibfnamefont {C.}~\bibnamefont {Dimopoulou}},
  \bibinfo {author} {\bibfnamefont {H.~C.}\ \bibnamefont {Hüther}}, \bibinfo
  {author} {\bibfnamefont {H.}~\bibnamefont {Liebermann}}, \bibinfo {author}
  {\bibfnamefont {J.}~\bibnamefont {Stadlmann}},\ and\ \bibinfo {author}
  {\bibfnamefont {R.}~\bibnamefont {Steinhagen}},\ }\bibfield  {title}
  {\bibinfo {title} {Recommissioning of {SIS18} {After} {FAIR} {Upgrades}},\
  }in\ \href@noop {} {\emph {\bibinfo {booktitle} {10th {Int}. {Particle}
  {Accelerator} {Conf}.({IPAC}'19), {Melbourne}, {Australia}, 19-24 {May}
  2019}}}\ (\bibinfo  {publisher} {JACOW Publishing, Geneva, Switzerland},\
  \bibinfo {year} {2019})\ pp.\ \bibinfo {pages} {932--935}\BibitemShut
  {NoStop}%
\bibitem [{\citenamefont {Herr}\ and\ \citenamefont
  {Forest}(2020)}]{Herr:Accelerator}%
  \BibitemOpen
  \bibfield  {author} {\bibinfo {author} {\bibfnamefont {W.}~\bibnamefont
  {Herr}}\ and\ \bibinfo {author} {\bibfnamefont {E.}~\bibnamefont {Forest}},\
  }\href {https://doi.org/10.1007/978-3-030-34245-6_3} {\bibinfo {title}
  {{Non-linear Dynamics in Accelerators}}} (\bibinfo {year} {2020})\BibitemShut
  {NoStop}%
\bibitem [{\citenamefont {Yoshida}(1990)}]{Yoshida:SymplecticIntegrators}%
  \BibitemOpen
  \bibfield  {author} {\bibinfo {author} {\bibfnamefont {H.}~\bibnamefont
  {Yoshida}},\ }\bibfield  {title} {\bibinfo {title} {{Construction of higher
  order symplectic integrators}},\ }\href
  {https://doi.org/10.1016/0375-9601(90)90092-3} {\bibfield  {journal}
  {\bibinfo  {journal} {Phys. Lett. A}\ }\textbf {\bibinfo {volume} {150}},\
  \bibinfo {pages} {262} (\bibinfo {year} {1990})}\BibitemShut {NoStop}%
\bibitem [{\citenamefont {Grote}\ and\ \citenamefont
  {Schmidt}(2003)}]{Grote:Mad-X}%
  \BibitemOpen
  \bibfield  {author} {\bibinfo {author} {\bibfnamefont {H.}~\bibnamefont
  {Grote}}\ and\ \bibinfo {author} {\bibfnamefont {F.}~\bibnamefont
  {Schmidt}},\ }\bibfield  {title} {\bibinfo {title} {{MAD-X: An upgrade from
  MAD8}},\ }\href@noop {} {\bibfield  {journal} {\bibinfo  {journal} {Conf.
  Proc. C}\ }\textbf {\bibinfo {volume} {030512}},\ \bibinfo {pages} {3497}
  (\bibinfo {year} {2003})}\BibitemShut {NoStop}%
\bibitem [{\citenamefont {Schwinzerl}\ \emph {et~al.}(2021)\citenamefont
  {Schwinzerl}, \citenamefont {Paraschou}, \citenamefont {De~Maria},
  \citenamefont {Iadarola}, \citenamefont {Oeftiger},\ and\ \citenamefont
  {Bartosik}}]{Schwinzerl:SixTrackLib}%
  \BibitemOpen
  \bibfield  {author} {\bibinfo {author} {\bibfnamefont {M.}~\bibnamefont
  {Schwinzerl}}, \bibinfo {author} {\bibfnamefont {K.}~\bibnamefont
  {Paraschou}}, \bibinfo {author} {\bibfnamefont {R.}~\bibnamefont {De~Maria}},
  \bibinfo {author} {\bibfnamefont {G.}~\bibnamefont {Iadarola}}, \bibinfo
  {author} {\bibfnamefont {A.}~\bibnamefont {Oeftiger}},\ and\ \bibinfo
  {author} {\bibfnamefont {H.}~\bibnamefont {Bartosik}},\ }\href@noop {} {\emph
  {\bibinfo {title} {Optimising and Extending a Single-particle Tracking
  Library for High Parallel Performance}}},\ \bibinfo {type} {Tech. Rep.}\
  (\bibinfo {year} {2021})\BibitemShut {NoStop}%
\bibitem [{\citenamefont {Qiang}(2023)}]{Qiang:DifferentiableSpaceCharge}%
  \BibitemOpen
  \bibfield  {author} {\bibinfo {author} {\bibfnamefont {J.}~\bibnamefont
  {Qiang}},\ }\bibfield  {title} {\bibinfo {title} {Differentiable
  self-consistent space-charge simulation for accelerator design},\ }\href@noop
  {} {\bibfield  {journal} {\bibinfo  {journal} {Physical Review Accelerators
  and Beams}\ }\textbf {\bibinfo {volume} {26}},\ \bibinfo {pages} {024601}
  (\bibinfo {year} {2023})}\BibitemShut {NoStop}%
\bibitem [{\citenamefont {C.~Duchi}\ \emph {et~al.}(2011)\citenamefont
  {C.~Duchi}, \citenamefont {Hazan},\ and\ \citenamefont {Singer}}]{Adagrad}%
  \BibitemOpen
  \bibfield  {author} {\bibinfo {author} {\bibfnamefont {J.}~\bibnamefont
  {C.~Duchi}}, \bibinfo {author} {\bibfnamefont {E.}~\bibnamefont {Hazan}},\
  and\ \bibinfo {author} {\bibfnamefont {Y.}~\bibnamefont {Singer}},\
  }\bibfield  {title} {\bibinfo {title} {Adaptive subgradient methods for
  online learning and stochastic optimization},\ }\href@noop {} {\bibfield
  {journal} {\bibinfo  {journal} {Journal of Machine Learning Research}\
  }\textbf {\bibinfo {volume} {12}},\ \bibinfo {pages} {2121} (\bibinfo {year}
  {2011})}\BibitemShut {NoStop}%
\bibitem [{\citenamefont {Zeiler}(2012)}]{Adadelta}%
  \BibitemOpen
  \bibfield  {author} {\bibinfo {author} {\bibfnamefont {M.~D.}\ \bibnamefont
  {Zeiler}},\ }\href@noop {} {\bibinfo {title} {Adadelta: An adaptive learning
  rate method}} (\bibinfo {year} {2012}),\ \Eprint
  {https://arxiv.org/abs/1212.5701} {arXiv:1212.5701 [cs.LG]} \BibitemShut
  {NoStop}%
\bibitem [{\citenamefont {Bartholomew-Biggs}\ \emph {et~al.}(2000)\citenamefont
  {Bartholomew-Biggs}, \citenamefont {Brown}, \citenamefont {Christianson},\
  and\ \citenamefont {Dixon}}]{Bartholomew:AutomaticDifferentiation}%
  \BibitemOpen
  \bibfield  {author} {\bibinfo {author} {\bibfnamefont {M.}~\bibnamefont
  {Bartholomew-Biggs}}, \bibinfo {author} {\bibfnamefont {S.}~\bibnamefont
  {Brown}}, \bibinfo {author} {\bibfnamefont {B.}~\bibnamefont
  {Christianson}},\ and\ \bibinfo {author} {\bibfnamefont {L.}~\bibnamefont
  {Dixon}},\ }\bibfield  {title} {\bibinfo {title} {Automatic differentiation
  of algorithms},\ }\href@noop {} {\bibfield  {journal} {\bibinfo  {journal}
  {Journal of Computational and Applied Mathematics}\ }\textbf {\bibinfo
  {volume} {124}},\ \bibinfo {pages} {171} (\bibinfo {year}
  {2000})}\BibitemShut {NoStop}%
\bibitem [{\citenamefont {Bezanson}\ \emph {et~al.}(2017)\citenamefont
  {Bezanson}, \citenamefont {Edelman}, \citenamefont {Karpinski},\ and\
  \citenamefont {Shah}}]{Bezanson:Julia}%
  \BibitemOpen
  \bibfield  {author} {\bibinfo {author} {\bibfnamefont {J.}~\bibnamefont
  {Bezanson}}, \bibinfo {author} {\bibfnamefont {A.}~\bibnamefont {Edelman}},
  \bibinfo {author} {\bibfnamefont {S.}~\bibnamefont {Karpinski}},\ and\
  \bibinfo {author} {\bibfnamefont {V.~B.}\ \bibnamefont {Shah}},\ }\bibfield
  {title} {\bibinfo {title} {Julia: A fresh approach to numerical computing},\
  }\href {https://doi.org/10.1137/141000671} {\bibfield  {journal} {\bibinfo
  {journal} {SIAM review}\ }\textbf {\bibinfo {volume} {59}},\ \bibinfo {pages}
  {65} (\bibinfo {year} {2017})}\BibitemShut {NoStop}%
\bibitem [{\citenamefont {Rackauckas}\ \emph {et~al.}(2021)\citenamefont
  {Rackauckas}, \citenamefont {Edelman}, \citenamefont {Fischer}, \citenamefont
  {Innes}, \citenamefont {Saba}, \citenamefont {Shah},\ and\ \citenamefont
  {Tebbutt}}]{Rackauckas:Zygote}%
  \BibitemOpen
  \bibfield  {author} {\bibinfo {author} {\bibfnamefont {C.}~\bibnamefont
  {Rackauckas}}, \bibinfo {author} {\bibfnamefont {A.}~\bibnamefont {Edelman}},
  \bibinfo {author} {\bibfnamefont {K.}~\bibnamefont {Fischer}}, \bibinfo
  {author} {\bibfnamefont {M.}~\bibnamefont {Innes}}, \bibinfo {author}
  {\bibfnamefont {E.}~\bibnamefont {Saba}}, \bibinfo {author} {\bibfnamefont
  {V.~B.}\ \bibnamefont {Shah}},\ and\ \bibinfo {author} {\bibfnamefont
  {W.}~\bibnamefont {Tebbutt}},\ }\bibfield  {title} {\bibinfo {title}
  {Generalized physics-informed learning through language-wide differentiable
  programming},\ }\href {https://dspace.mit.edu/handle/1721.1/137320}
  {\bibfield  {journal} {\bibinfo  {journal} {MIT web domain}\ } (\bibinfo
  {year} {2021})},\ \bibinfo {note} {accepted:
  2021-11-04T11:58:19Z}\BibitemShut {NoStop}%
\bibitem [{\citenamefont {Innes}(2018)}]{Innes:FluxML}%
  \BibitemOpen
  \bibfield  {author} {\bibinfo {author} {\bibfnamefont {M.}~\bibnamefont
  {Innes}},\ }\bibfield  {title} {\bibinfo {title} {Flux: Elegant machine
  learning with julia},\ }\bibfield  {journal} {\bibinfo  {journal} {Journal of
  Open Source Software}\ }\href {https://doi.org/10.21105/joss.00602}
  {10.21105/joss.00602} (\bibinfo {year} {2018})\BibitemShut {NoStop}%
\bibitem [{\citenamefont {Bergstra}\ \emph {et~al.}(2011)\citenamefont
  {Bergstra}, \citenamefont {Bardenet}, \citenamefont {Bengio},\ and\
  \citenamefont {K{\'e}gl}}]{Bergstra:TPE}%
  \BibitemOpen
  \bibfield  {author} {\bibinfo {author} {\bibfnamefont {J.}~\bibnamefont
  {Bergstra}}, \bibinfo {author} {\bibfnamefont {R.}~\bibnamefont {Bardenet}},
  \bibinfo {author} {\bibfnamefont {Y.}~\bibnamefont {Bengio}},\ and\ \bibinfo
  {author} {\bibfnamefont {B.}~\bibnamefont {K{\'e}gl}},\ }\bibfield  {title}
  {\bibinfo {title} {Algorithms for hyper-parameter optimization},\ }\href@noop
  {} {\bibfield  {journal} {\bibinfo  {journal} {Advances in neural information
  processing systems}\ }\textbf {\bibinfo {volume} {24}} (\bibinfo {year}
  {2011})}\BibitemShut {NoStop}%
\bibitem [{\citenamefont {Akiba}\ \emph {et~al.}(2019)\citenamefont {Akiba},
  \citenamefont {Sano}, \citenamefont {Yanase}, \citenamefont {Ohta},\ and\
  \citenamefont {Koyama}}]{Akiba:Optuna}%
  \BibitemOpen
  \bibfield  {author} {\bibinfo {author} {\bibfnamefont {T.}~\bibnamefont
  {Akiba}}, \bibinfo {author} {\bibfnamefont {S.}~\bibnamefont {Sano}},
  \bibinfo {author} {\bibfnamefont {T.}~\bibnamefont {Yanase}}, \bibinfo
  {author} {\bibfnamefont {T.}~\bibnamefont {Ohta}},\ and\ \bibinfo {author}
  {\bibfnamefont {M.}~\bibnamefont {Koyama}},\ }\bibfield  {title} {\bibinfo
  {title} {Optuna: A next-generation hyperparameter optimization framework},\
  }in\ \href@noop {} {\emph {\bibinfo {booktitle} {Proceedings of the 25rd
  {ACM} {SIGKDD} International Conference on Knowledge Discovery and Data
  Mining}}}\ (\bibinfo {year} {2019})\BibitemShut {NoStop}%
\bibitem [{\citenamefont {Barth}\ \emph {et~al.}(2015)\citenamefont {Barth},
  \citenamefont {Adonin}, \citenamefont {Appel}, \citenamefont {Gerhard},
  \citenamefont {Heilmann}, \citenamefont {Heymach}, \citenamefont {Hollinger},
  \citenamefont {Vinzenz}, \citenamefont {Vormann},\ and\ \citenamefont
  {Yaramyshev}}]{Barth:Unilac}%
  \BibitemOpen
  \bibfield  {author} {\bibinfo {author} {\bibfnamefont {W.}~\bibnamefont
  {Barth}}, \bibinfo {author} {\bibfnamefont {A.}~\bibnamefont {Adonin}},
  \bibinfo {author} {\bibfnamefont {S.}~\bibnamefont {Appel}}, \bibinfo
  {author} {\bibfnamefont {P.}~\bibnamefont {Gerhard}}, \bibinfo {author}
  {\bibfnamefont {M.}~\bibnamefont {Heilmann}}, \bibinfo {author}
  {\bibfnamefont {F.}~\bibnamefont {Heymach}}, \bibinfo {author} {\bibfnamefont
  {R.}~\bibnamefont {Hollinger}}, \bibinfo {author} {\bibfnamefont
  {W.}~\bibnamefont {Vinzenz}}, \bibinfo {author} {\bibfnamefont
  {H.}~\bibnamefont {Vormann}},\ and\ \bibinfo {author} {\bibfnamefont
  {S.}~\bibnamefont {Yaramyshev}},\ }\bibfield  {title} {\bibinfo {title}
  {Heavy ion linac as a high current proton beam injector},\ }\href@noop {}
  {\bibfield  {journal} {\bibinfo  {journal} {Physical Review Special
  Topics-Accelerators and Beams}\ }\textbf {\bibinfo {volume} {18}},\ \bibinfo
  {pages} {050102} (\bibinfo {year} {2015})},\ \bibinfo {note} {publisher:
  APS}\BibitemShut {NoStop}%
\bibitem [{\citenamefont {Appel}\ and\ \citenamefont
  {Boine-Frankenheim}(2012)}]{Appel:Microbunch}%
  \BibitemOpen
  \bibfield  {author} {\bibinfo {author} {\bibfnamefont {S.}~\bibnamefont
  {Appel}}\ and\ \bibinfo {author} {\bibfnamefont {O.}~\bibnamefont
  {Boine-Frankenheim}},\ }\bibfield  {title} {\bibinfo {title} {{Microbunch
  dynamics and multistream instability in a heavy-ion synchrotron}},\ }\href
  {https://doi.org/10.1103/PhysRevSTAB.15.054201} {\bibfield  {journal}
  {\bibinfo  {journal} {Phys. Rev. Accel. Beams}\ }\textbf {\bibinfo {volume}
  {15}},\ \bibinfo {pages} {054201} (\bibinfo {year} {2012})}\BibitemShut
  {NoStop}%
\bibitem [{\citenamefont {Franczak}(1987)}]{Franczak:SIS18_parameterList}%
  \BibitemOpen
  \bibfield  {author} {\bibinfo {author} {\bibfnamefont {B.}~\bibnamefont
  {Franczak}},\ }\bibfield  {title} {\bibinfo {title} {Sis parameter list},\
  }\href@noop {} {\bibfield  {journal} {\bibinfo  {journal} {Techn. Ber.
  GSI-SIS-TN/87-13. Gesellschaft f{\"u}r Schwerionenforschung}\ } (\bibinfo
  {year} {1987})}\BibitemShut {NoStop}%
\end{thebibliography}%

\end{document}